\documentclass[technote,10pt]{IEEEtran}
\usepackage{cite}
\usepackage{graphicx}
\usepackage{picinpar}
\usepackage[cmex10]{amsmath}
\usepackage{amsmath,amsfonts,amssymb}
\usepackage{subfigure}
\usepackage{amsthm}
\usepackage{algorithm}
\usepackage{algorithmic}
\usepackage{stfloats}
\usepackage{bm}
\usepackage{color}
\usepackage{booktabs}
\usepackage{makecell}
\usepackage{mathrsfs}
\usepackage{stfloats}
\usepackage{url}
\usepackage{epstopdf}
\DeclareMathAlphabet{\mathbbold}{U}{bbold}{m}{n}
\begin{document}
\pdfoptionpdfminorversion=6
\newtheorem{lemma}{Lemma}
\newtheorem{corol}{Corollary}
\newtheorem{theorem}{Theorem}
\newtheorem{proposition}{Proposition}
\newtheorem{definition}{Definition}
\newcommand{\e}{\begin{equation}}
\newcommand{\ee}{\end{equation}}
\newcommand{\eqn}{\begin{eqnarray}}
\newcommand{\eeqn}{\end{eqnarray}}
\renewcommand{\algorithmicrequire}{ \textbf{Input:}} 
\renewcommand{\algorithmicensure}{ \textbf{Output:}} 
\title{Deep Denoising Neural Network Assisted Compressive Channel Estimation for mmWave Intelligent Reflecting Surfaces}

\author{Shicong Liu, Zhen Gao, Jun Zhang, Marco Di Renzo,~\IEEEmembership{Fellow,~IEEE}, Mohamed-Slim Alouini,~\IEEEmembership{Fellow,~IEEE} 
\vspace*{-3.0mm}
\thanks{S. Liu, Z. Gao and J. Zhang are with the Advanced Research Institute of Multidisciplinary Science (ARIMS) and School of Information and Electronics,
Beijing Institute of Technology (BIT), Beijing 100081, China.}
\thanks{M. Di Renzo is with the Laboratoire des Signaux et Syst\`emes, CNRS, CentraleSup\'elec, Univ Paris Sud, Universit\'e Paris-Saclay, 91192 Paris, France.
}
\thanks{M.-S. Alouini is with the Electrical Engineering Program, Division of
Physical Sciences and Engineering, King Abdullah University of Science and
Technology, Thuwal 23955, Saudi Arabia.}
}
\maketitle
\begin{abstract}
Integrating large intelligent reflecting surfaces (IRS) into millimeter-wave (mmWave) massive multi-input-multi-ouput (MIMO) has been a promising approach for improved coverage and throughput. Most existing work assumes the ideal channel estimation, which can be challenging due to the high-dimensional cascaded MIMO channels and passive reflecting elements.
Therefore, this paper proposes a deep denoising neural network assisted compressive channel estimation for mmWave IRS systems to reduce the training overhead. Specifically, we first introduce a hybrid passive/active IRS architecture, where very few receive chains are employed to estimate the uplink user-to-IRS channels. At the channel training stage, only a small proportion of elements will be successively activated to sound the partial channels. Moreover, the complete channel matrix can be reconstructed from the limited measurements based on compressive sensing, whereby the common sparsity of angular domain mmWave MIMO channels among different subcarriers is leveraged for improved accuracy. Besides, a complex-valued denoising convolution neural network (CV-DnCNN) is further proposed for enhanced performance. Simulation results demonstrate the superiority of the proposed solution over state-of-the-art solutions.
\end{abstract}

\vspace*{-1mm}
\vspace*{-2.0mm}
\IEEEpeerreviewmaketitle
\section{Introduction}
Millimeter-wave (mmWave) massive multiple-input multiple-output (MIMO) system with intelligent reflecting surfaces (IRS) is envisioned to be a promising technology for future beyond 5G/6G wireless communication networks to substantially improve the link quality and reduce the blockage probability\cite{ref:zhangrui2,3fellow}. IRS is a kind of near-passive array whose elements are able to attach independently controllable phase shifts to the incident electromagnetic waves\cite{EuRIS}. With very few radio frequency (RF) chains connected and massive near-passive elements, adopting IRS would be an energy-saving and low-cost solution. Leveraging the controllable reflection property of large IRS with passive elements, the mmWave massive MIMO base stations (BS) can serve the user equipments (UEs) whose direct links from the BS are completely blocked by obstacles, so that the effective coverage of mmWave cellular networks can be enhanced by some sophisticated beamforming techniques \cite{ref:zhangrui2,3fellow,ref:jointDesign}.

However, these substantial performance enhancements benefited from IRS rely on perfect channel state information (CSI), and the accurate channel estimation with reduced overhead still remains a challenge for IRS with a large number of passive elements. On the one hand, the cascaded MIMO channel between the BS and UE via IRS can be extremely high-dimensional. On the other hand, the passive feature of IRS makes it rather difficult to respectively estimate IRS-UE and IRS-BS channels. Prior work mainly focuses on the design of reflection matrix by assuming the BS and IRS to perfectly know the CSI\cite{ref:jointDesign,fjprecoding,lyhprecoding}, whose estimation remains a challenge. Against this background, several channel estimation approaches have been proposed in\cite{icassp,CascadedCE,rzCE,fjce,aa}. In \cite{icassp}, by assuming an element-by-element ON/OFF-based reflection pattern, the authors proposed a least squares (LS) based estimation method. Furthermore, by exploiting the low rank property of MIMO channels and designing a random ON/OFF reflection pattern, a sparse matrix factorization based cascaded channel estimation solution was proposed in \cite{CascadedCE} with much reduced pilot overhead. Considering the sparse representation of the cascaded channel, the authors in \cite{fjce} proposed a compressive sensing (CS) based channel estimation method, which can estimate the cascaded channel simultaneously with substantial training overhead reduction.
Nevertheless, solutions in \cite{icassp}, \cite{CascadedCE} and \cite{fjce} mainly consider a frequency-flat communication system by assuming narrow-band channels.
Therefore, to fully exploit the full reflection of the IRS during the channel estimation phase, the authors in \cite{rzCE} applied the LS to jointly estimate the BS-UE and BS-IRS-UE channels in orthogonal frequency division multiplexing (OFDM) systems,
however, they only consider all UEs and BS to be equipped with a single antenna, as the prohibitive pilot overhead or channel dimension may be extremely high when the numbers of antennas become large.
Besides, two channel estimation schemes respectively based on compressive sensing and deep learning (DL) were proposed in \cite{aa} and \cite{aaa}, whereby the angular-domain channel sparsity was utilized for reduced pilot overhead. The proposed hybrid passive/active IRS was efficient for channel estimation, however, the adopted 2D-discrete Fourier transform (DFT) matrix in CS formulation may suffer from a performance loss due to the unnegligible power leakage effect \cite{Tcomm}, and the structured channel sparsity among different subcarriers were not fully exploited.

In this paper, we propose a deep denoising neural network assisted CS broadband channel estimation for mmWave IRS systems to reduce the training overhead\footnote{The python code can be found in https://github.com/psycholsc/complex-DnCNN}. To be more specific,
we firstly introduce the hybrid passive/active IRS architecture\cite{aa}, where very few receive RF chains were employed by trading off the channel estimation performance with power consumption as well as hardware complexity. On this basis, we propose a CS-based broadband channel estimation solution, where the simultaneous orthogonal match pursuit (SOMP) algorithm\cite{WC} is adopted to jointly estimate the channels of multiple subcarriers by exploiting the multiple measurement vector (MMV) property.
Besides, by leveraging the correlation of angular-delay domain MIMO channel matrix, developed from the denoising convolution neural network (DnCNN)\cite{DnCNN}, a complex-valued DnCNN (CV-DnCNN) is proposed to further enhance the estimation accuracy. 
Our contributions are summarized as follows:
\begin{itemize}
  \item On the basis of the hybrid passive/active IRS architecture, we propose a corresponding multi-carrier pilot transmission scheme, which can be applied to IRS-assisted mmWave hybrid MIMO systems.
  \item The proposed CS channel estimation approach leverages the angular-domain common sparsity of mmWave MIMO channels over different subcarriers and adopts a redundant dictionary for improved accuracy.
  \item We develop a CV-DnCNN tailored to complex-valued channels for further enhanced estimation accuracy. The proposed CV-DnCNN can work in various signal-to-noise-ratios (SNRs) and numbers of multipath components (MPCs) even it is trained at a certain SNR or a certain number of MPCs.
\end{itemize}



\textit{Notation}:
Scalar variables are denoted by normal-face letters, while boldface lower and upper-case symbols denote column vectors and matrices, respectively.
$\mathbb{C}$ and $\mathbb{Z}$ are the sets of complex numbers and integers, respectively.
Superscripts $ (\cdot)^{*},(\cdot)^{T}, (\cdot)^{H} $ denote the conjugate, transpose, conjugate transpose operators, respectively.
$ {\|\bf{A}\|}_{F} $ is the Frobenius norm of $\bf{A}$, while $\otimes$ and $\circledast$ denote the Kronecker product and convolution operation, respectively. Further, we use ${\rm diag}(\cdot)$ and ${\rm vec}(\cdot)$ to denote the diagonalization operation (from vectors to the diagonal matrices) and vectorization operation, respectively. $\mathfrak{Re}(\cdot)$ and $\mathfrak{Im}(\cdot)$ denote the real part and imaginary part of complex numbers. Key model-related notations are listed in Table \ref{notation}.
\renewcommand\thetable{\Roman{table}}
\begin{table}[h]
  \centering
  \caption{Model-related notations}\label{notation}
  \begin{tabular}{c|c}
  \Xhline{1.2pt}
  {\bf{Notation}} & {\bf{Definition}}\\
  \Xhline{1.2pt}
  ${\bf{W}_{\rm{RF}}}\in \mathbb{C}^{ N^{\rm{ant}}_{\rm{BS}} \times N_{\rm{BS}}^{\rm{RF}} }$ & BS analog combiner \\
  \hline
  ${{\bf{W}}_{{\rm{BB}},k} }\in \mathbb{C}^{N_{\rm{BS}}^{\rm{RF}}\times N_{\rm{BS}}^{\rm{S}}}$ & BS digital combiner \\
  \hline
  ${{\bf{F}}_{\rm{RF}}}\in \mathbb{C}^{N^{\rm{ant}}_{\rm{UE}} \times N_{\rm{UE}}^{\rm{RF}}}$ & UE analog precoder \\
  \hline
  ${{\bf{F}}_{{\rm{BB}},k}}\in\mathbb{C}^{N_{\rm{UE}}^{\rm{RF}}\times N_{\rm{UE}}^{\rm{S}}}$ & UE digital precoder  \\
  \hline
  ${{\bf{H}}_{1,k}}\in\mathbb{C}^{N^{\rm{ant}}_{\rm{BS}} \times N^{\rm{ant}}_{\rm{IRS}}}$ & BS-IRS channel  \\
  \hline
  ${{\bf{H}}_{2,k}}\in\mathbb{C}^{N^{\rm{ant}}_{\rm{IRS}} \times N^{\rm{ant}}_{\rm{UE}}}$ & IRS-UE channel  \\
  \hline
  ${\bf{\Theta}}\in\mathbb{C}^{N^{\rm{ant}}_{\rm{IRS}} \times N^{\rm{ant}}_{\rm{IRS}}}$ & Diagonal IRS reflecting matrix  \\
  \hline
  ${\bf{x}}_k\in\mathbb{C}^{N^{\rm S}_{\rm UE}}$ & Data signal vector   \\
  \hline
  ${\bf{r}}_k\in\mathbb{C}^{N^{\rm S}_{\rm BS}}$  & Received signal vector \\
  \hline
  ${\bf{w}}_k\in\mathbb{C}^{N^{\rm ant}_{\rm BS}}$ & Noise signal vector\\
  \hline
  ${\bf{W}}^{b}_{\rm{AS}} \in \mathbb{Z}^{N^{\rm{S}}_{\rm{IRS}} \times N^{\rm{ant}}_{\rm{IRS}}}$ & IRS antenna selection matrix\\
  \Xhline{1.2pt}
\end{tabular}
\end{table}
\section{System Model}

\begin{figure}[t]
\centering
\includegraphics[scale=0.5]{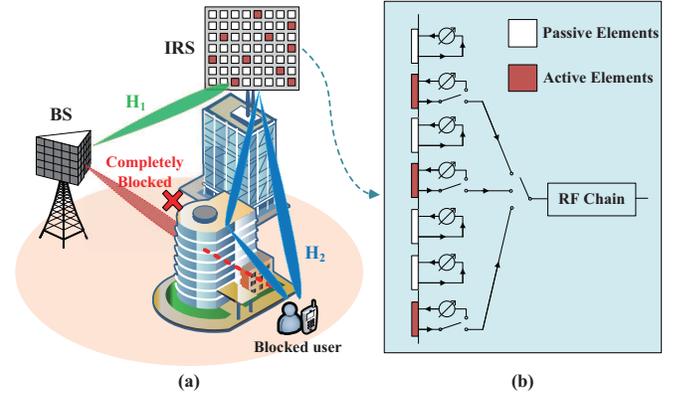}
\caption{IRS-aided mmWave massive MIMO systems: (a) IRS can improve mmWave coverage for UEs that have no effective link from the BS; (b) Introduced hybrid passive/active IRS architecture based on antenna switch network.}
\label{fig:commSys}
\end{figure}

Consider an IRS-aided mmWave massive MIMO system as shown in Fig. \ref{fig:commSys}(a). OFDM with $K$ subcarriers is adopted to combat the time-dispersive channels, and the IRS is utilized to reflect the incident signals with controllable phases. We focus on the UEs whose direct links to the mmWave BS are completely blocked by obstacles, e.g., the buildings as shown in Fig. \ref{fig:commSys}(a). So the uplink received signal at the BS associated with the $k$-th ($1\!\leq\! k\!\leq\! K$) subcarrier, denoted by ${\bf{r}}_k \in \mathbb{C}^{N_{\rm{BS}}^{\rm{S}} \times 1}$, can be expressed as
\begin{equation}
{{\bf r}}_k\!=\!{\left({{\bf{W}}_{\rm{RF}}}\!{\bf{W}}_{{\rm{BB}},k}\!\right)}^H\!\!  \left(  {{\bf{H}}_{1,k}} {{\bf{\Theta}}}  {{\bf{H}}_{2,k}} {{\bf{F}}_{{\rm{RF}}}}{{\bf{F}}_{{\rm{BB}},k}}{{\bf x}}_k\!+\!{{{\bf w}}}_k\right),
\label{systemModel1}
\end{equation}
where the notations are described in Table \ref{notation}. For the hybrid MIMO employed at the transceiver, $N^{\rm{ant}}_{\rm BS}$ ($N^{\rm{ant}}_{\rm UE}$), $N^{\rm{RF}}_{\rm BS}$ ($N^{\rm{RF}}_{\rm UE}$), and $N^{\rm{S}}_{\rm BS}$ ($N^{\rm{S}}_{\rm UE}$) denote the numbers of antennas, RF chains, and data streams at BS (UEs), respectively, and  $N^{\rm{S}}_{\rm BS}$ = $N^{\rm{S}}_{\rm UE}$ is assumed for single-user scenario. Note that the proposed channel estimation scheme can be extended to multi-user scenario by using mutually orthogonal time or code domain resources to transmit different UEs' pilot signals\cite{QTcomm}. In this way, the receiver can easily distinguish different UEs' pilot signals and further separately estimate their channels.As shown in Fig. \ref{fig:commSys}, IRS is introduced to serve the UEs located in the coverage ``dead zones'', where the direct links between BS and UE does not exist. To ensure the service quality, IRSs are usually deployed in the line-of-sight (LoS) areas of BSs. Since the locations of BSs and IRSs are fixed once they are deployed, the LoS AoA/AoD connecting BS and IRS would remain unchanged for a long time. Therefore, the channel between BS and IRS can be easily estimated by one-step least squares with previously calibrated accurate angles. Based on the above analysis, we would only focus on estimation of changeable IRS-UE channels.

\section{Proposed Channel Estimation Technique}
In this section, we propose a CV-DnCNN assisted CS channel estimation technique for mmWave IRS systems, where the UE-IRS channels can be estimated with reduced training overhead. Specifically, we first introduce a hybrid passive/active IRS architecture and proposed an associated uplink multi-carrier channel estimation scheme based on pilot signals. Moreover, from the very few measurements collected at the IRS, CS and DL are utilized to reconstruct the high-dimensional UE-IRS channels with improved estimation accuracy.
\subsection{Pilot Training with Hybrid Passive/Active IRS Architecture}
We consider the IRS adopts the practical uniform planar array (UPA) consisting of a large number of passive elements and very few active elements, which are respectively marked in white color and red color as shown in Fig. \ref{fig:commSys}(a). The passive elements only attach phase shifts to the incident signals and then reflect them, while the active elements can be sequentially connected to the $N^{\rm RF}_{\rm IRS}$ receive RF chains via an antenna switching network as shown in Fig. \ref{fig:commSys}(b). During the pilot training phase, every $N^{\rm{RF}}_{\rm IRS}$ active elements are activated so that we can obtain $N^{\rm{RF}}_{\rm IRS}$ measurements in each time slot. During the data transmission phase, the active elements will work as the passive ones.
The uplink pilot training phase consists of $B$ time slots (also $B$ OFDM symbols), and the received signal ${\bf y}_k^b \in \mathbb{C}^{N^{\rm S}_{\rm IRS}}$ at the IRS associated with the $k$-th subcarrier and $b$-th time slot can be written as
\begin{equation}
{\bf{y}}^{b}_{k}={{\bf{W}}^b_{\rm{AS}} }  {{\bf{H}}_{k}} {{\bf{F}}^{b}_{{\rm RF}}}{{\bf{F}}^{b}_{{\rm{BB}},k}}{\bf{s}}^{b}_{k}+{\bf{n}}^{b}_{k} ,
\label{mk}
\end{equation}
where each row of the antenna selection matrix ${\bf{W}}^{b}_{\rm{AS}}$ is filled with ($N_{\rm IRS}^{\rm ant}-1$) $0$ entries and only one $1$ entry (i.e., $N_{\rm IRS}^{\rm S}$ stacked one-hot vectors), analog and digital precoders in the pilot training stage ${{\bf{F}}^{b}_{{\rm RF}}}$ and ${{\bf{F}}^{b}_{{{\rm BB},k}}}$ at the UE are pre-defined and known by the receiver\cite{QTcomm}, and here we consider ${{\bf{F}}^{b}_{{{\rm BB},k}}}$ to be an identity matrix and ${{\bf{F}}^{b}_{{{\rm RF},k}}}$ to meet the constant modulus constraint with its elements' phases following the mutually independent uniform distribution $\mathcal{U}[0,2\pi]$. ${\bf s}_k^b\in\mathbb{C}^{N_{\rm{UE}}^{\rm{S}}}$ represents the pilot vector in the $b$-th time slot associated with the $k$-th subcarrier, and it can be a full-one vector. ${\bf{n}}^{b}_{k}\in\mathcal{CN}(0,\sigma_n^2 {\bf{I}}_{N^{\rm S}_{\rm{IRS}}})$ is the additive white Gaussian noise vector, and here ${\bf H}_k = {\bf H}_{2,k}$ in Eq. \eqref{systemModel1}. More specifically, ${{\bf{H}}_{k}}$ can be expressed as
\begin{equation}
\label{freqChannel}
{{\bf{H}}_{k}}=\sum_{d=0}^{K-1} {{\bf{C}}_d} e^{-j\frac{2\pi k}{K}d},
\end{equation}
where ${\bf{C}}_d$ is the mmWave MIMO discrete channel impulse response in the discrete delay domain, and it can be modeled as the geometric channel model\cite{aa,QPS}
\begin{equation}
{\bf{C}}_d\!=\!\sqrt{\frac{N^{\rm{ant}}_{\rm{UE}} N^{\rm{ant}}_{\rm{IRS}}}{L}}\sum_{\ell =1}^L \alpha_\ell p\left( dT_s\!-\!\tau_\ell \right){\bf{a}}_{\rm{R}}(\phi_\ell,\psi_\ell){\bf{a}}_{\rm{T}}^*(\phi_{\ell}{'},\psi_{\ell}{'}),
\label{gm}
\end{equation}
where $L$, $T_{\rm{s}}$, and $p\left(\tau \right)$ denote the number of MPCs, sampling period, and pulse shaping filter, respectively. $\phi$ ($\phi{'}$) and $\psi$ ($\psi{'}$) are respectively the azimuth and elevation angle at IRS (UEs), $\alpha_\ell$ and $\tau_\ell$ denote the path gain and the discrete delay of the $\ell$-th MPC, respectively. Note that the path gain $\alpha_\ell$ depends on several parameters such as (i) distance $r_d$ between UEs and IRS, (ii) carrier frequency $f_c$, (iii) transmit power $P_T$, (iv) antenna gain, and so on. The path loss can be high in mmWave frequency, for example with $r_d=300$m, $P_T=20$dBm, $f_c=28$GHz scenario, the equivalent receive SNR can be as low as $-10$dB and as high as 20dB, according to the measurement model in \cite{snrrange}. Therefore, in our simulations, SNR range would start from $-10$dB and end at $20$dB. ${\bf{a}}_{\rm{R}}(\phi,\psi)$ and ${\bf{a}}_{\rm{T}}(\phi{'},\psi{'})$ are the UPA's steering vectors for the IRS and UE, respectively, and share the similar expression. ${\bf{a}}_{\rm{R}}(\phi,\psi)$ can be expressed as
\begin{align}
\!{{\bf{a}}}_{\rm R}  (\phi_\ell,\psi_\ell) = \frac{1}{\sqrt {N^{\rm{ant}}_{\rm{IRS}}} }[ 1\!,\!\cdots \!,\!{e^{j\frac{{2\pi }}{\lambda }d\left( {n\sin \left( \phi_\ell   \right) \cos\left( \psi_\ell \right) + m\sin \left( \psi_\ell  \right)} \right)}}\!,\!\notag\\ \cdots \!, {e^{j\frac{{2\pi }}{\lambda }d ( {\left( {N_{\rm IRS}^{\rm{H}}\!-\!1} \right)\sin ( \phi_\ell )\cos ( \psi_\ell   ) + ( {N_{\rm IRS}^{\rm{W}}\!-\!1} )\sin ( \psi_\ell   )} )}}]^T.\notag
\end{align}
Rewrite the channel in a compact way, we have ${\bf{C}}_d={\bf{A}}_{\rm{R}}{\bf{\tilde C}}_d {\bf{A}}_{\rm{T}}^{{H}}$, where
\begin{align}
{\bf{\tilde C}}_d \!=\! \sqrt{\frac{N^{\rm{ant}}_{\rm{UE}} N^{\rm{ant}}_{\rm{IRS}}}{L}}\! {\rm diag} (\left[ \alpha_1 p( dT_s-\tau_1 ),\!\cdots\!,\alpha_L p( dT_s-\tau_L ) \right]),\notag
\end{align}
and ${\bf{A}}_{\rm{R}}= [{\bf{a}}_{\rm{R}}(\phi_1,\psi_1),\cdots,{\bf{a}}_{\rm{R}}(\phi_{{L}},\psi_{{L}} )] \in \mathbb{C}^{N^{\rm{ant}}_{\rm{IRS}}\times L }$, ${\bf{A}}_{\rm{T}}=$ $[{\bf{a}}_{\rm{T}}(\phi{'}_1,\psi{'}_1),\cdots,{\bf{a}}_{\rm{T}}(\phi{'}_{{L}},\psi{'}_{{L}} )] \in \mathbb{C}^{N^{\rm{ant}}_{\rm{UE}}\times L }$ are matrices that contain the steering vectors of AoAs and AoDs, respectively.
By applying the vectorization operation on \eqref{mk}, we further obtain
\begin{align}
{\bf{y}}^{b}_{k}={\rm vec}({\bf{y}}^{b}_{k})&=\left({{\bf{F}}^{b}}{\bf{s}}^{b}\right)^T\!\otimes\! \left({\bf W}_{\rm AS}^b\right) \!{\rm vec}({{\bf{H}}_{{{k}}}})\!+\!{\bf{{n}}}^{b}_{k}\notag\\&={\bf \Phi}^{b}{{\bf{h}}_{{{k}}}}+{\bf{{n}}}^{b}_{k},
\label{pilot}
\end{align}
where ${\bf \Phi}^b=\left({{\bf{F}}^{b}}{\bf{s}}^{b}\right)^T\otimes \left({{\bf{W}}_{\rm AS}^b }\right)$ is the measurement matrix of $b$-th time slot, ${\bf F}^b={\bf F}_{\rm RF}^b{\bf F}_{\rm BB}^b$ and ${\bf W}_{\rm AS}^b$ are respectively the frequency-flat precoders and antenna selectors, and ${\bf{h}}_{k}={\rm{vec}}({\bf{H}}_{k})$ is the vectorized channel. Note that the frequency-flat assumption is convenient for calculation, while this assumption would potentially result in high peak-to-average-power-ratio (PAPR). We can introduce a pseudo-random scrambling code to relax that assumption, which can be seen in \cite{Tcomm} for details. It is obvious that the number of measurements in each time slot equals the number of RF chains $N_{\rm IRS}^{\rm RF}$, which implies that given the total number of measurements $M=BN_{\rm IRS}^{\rm RF}$, the training overhead $B$ can be reduced by increasing $N_{\rm IRS}^{\rm RF}$, at the cost of increased power consumption and hardware requirement. Therefore, we can adopt only 1 RF chain for energy-saving and low cost purposes. After $M=BN_{\rm IRS}^{\rm RF}$ active elements are successively activated in $B$ time slots, we obtain an aggregate observation given by
\begin{align}
{\bf y}_k&\!=\!\left[ ({\bf y}_k^1)^T,({\bf y}_k^2)^T,\cdots ,({\bf y}_k^{B})^T \right]^T\notag\\&\!=\![({\bf{\Phi}}^1)^T,\cdots,({\bf{\Phi}}^{B})^T]^T {\bf {h}}_k\! +\!{\bf{n}}_{{{k}}}\!=\!{\bf{\Phi}} {\bf{{h}}}_k\!+\!{\bf{n}}_{k},
\label{Y}
\end{align}
where ${\bf \Phi}=[({\bf \Phi}^1)^T,({\bf \Phi}^2)^T,\cdots,({\bf \Phi}^B)^T]^T \in \mathbb{C}^{ M \times N_{\rm IRS}^{\rm ant} N_{\rm UE}^{\rm ant} }$ represents the aggregate measurement matrix, and ${\bf n}_k=[({\bf n}^1_k)^T,({\bf n}^2_k)^T,\cdots,({\bf n}^{B}_k)^T]^T$ is the stacked noise vector.
\subsection{CS-Based UE-IRS Channel Reconstruction}
Accurately estimating the high-dimensional channels $\{{\bf h}_k\}_{k=1}^{K}$ from \eqref{Y}
usually requires $M\geq N_{\rm IRS}^{\rm ant}N_{\rm UE}^{\rm ant}$. Fortunately, thanks to the angular-domain sparsity feature of mmWave MIMO channels, we can apply CS theory to this estimation problem for reduced pilot overhead\cite{WC}. Nevertheless, considering the limited resolution in angular domain of discrete dictionary, the power leakage caused by the mismatch between continuous AoAs/AoDs and discrete dictionary grids may weaken the sparsity. To mitigate this phenomenon, we design a redundant dictionary and rewrite the channel ${\bf H}_k$ in \eqref{freqChannel} represented in the redundant dictionary as
\begin{equation}
{\bf{H}}_k = {\bf{A}}_{\rm{R}}^{\rm{D}}  {{{\bf{\tilde H}}}_k}  \left({\bf{A}}_{\rm{T}}^{\rm{D}}\right)^{\rm{H}}+{\bf{\bar{N}}},
\label{dict}
\end{equation}
where ${\bf{A}}_{\rm{R}}^{\rm{D}}$ and ${\bf{A}}_{\rm{T}}^{\rm{D}}$ share the same form as

\begin{equation}
\begin{aligned}
{\bf{A}}_{\rm{R}}^{\rm{D}}=[{\bf{a}}_{\rm{R}}(\phi_1,\psi_1),\cdots,{\bf{a}}_{\rm{R}}(\phi_1,\psi_{\beta N^{\rm H}_{\rm IRS}}),\\ \cdots,{\bf{a}}_{\rm{R}}(\phi_{\beta N^{\rm{W}}_{\rm IRS}},\psi_{\beta N^{\rm{H}}_{\rm IRS}})],
\end{aligned}
\end{equation}
$\phi_i$, $\psi_i$ are the uniformly selected angle grids in the range of $[-\pi/2,\pi/2]$, and they can be given by $\phi_i=-\pi/2+{i\pi}/{\beta N_{\rm IRS}^{\rm W}}$ ($i=1,2,\cdots,\beta N_{\rm IRS}^{\rm W}$) and $\psi_i=-\pi/2+{i\pi}/{\beta N_{\rm IRS}^{\rm H}}$ ($i=1,2,\cdots,\beta N_{\rm IRS}^{\rm H}$). ${\bf{\bar{N}}}$ denotes the quantization error matrix treated as a random noise, $\beta$ is the oversampling rate, and ${{{\bf{\tilde H}}}_k}$ is an approximate $\tilde L$-sparse angular domain channel represented by the redundant dictionary, whose sparsity can be weakened by power leakage\cite{Tcomm}.
Therefore, we set an oversampling factor $\beta >1$ to mitigate the leakage. Note that when $\beta=1$, the dictionary matrix is equivalent to the Kronecker product of two DFT matrices. Substitute (\ref{dict}) into (\ref{Y}), we have
\begin{equation}
\begin{aligned}
\!{\bf{y}}_k\!=\!{\bf{\Phi}} {\rm{vec}}\left({\bf{A}}_{\rm{R}}^{\rm{D}}  {{{\bf{\tilde H}}}_k}  \left({\bf{A}}_{\rm{T}}^{\rm{D}}\right)^{\rm{H}}\!+\!{\bf{\bar{N}}}\right) +{\bf{n}}_{{{k}}}={\bf{\Phi \Psi}} {\bf{\tilde{h}}}_k+{\bf{n}}_{\rm{E}},
\label{measurement}
\end{aligned}
\end{equation}
where ${\bf{\Psi}}=\left({\bf{A}}_{\rm{T}}^{\rm{D}}\right)^{*}\otimes {\bf{A}}_{\rm{R}}^{\rm{D}}$ denotes the redundant dictionary matrix, ${\bf{\tilde{h}}}_k={\rm{vec}}({{{\bf{\tilde H}}}_k})$ denotes the the sparse formulation under the basis of redundant dictionary, and ${\bf{n}}_{\rm{E}}={\bf \Phi}{\rm{vec}}({\bf{\bar{N}}})+{\bf{n}}_{k}$ denotes the effective noise. Since the spatial propagation characteristics of the channels within the system bandwidth are almost unchanged, the subchannels associated with different subcarriers share very similar scatterers in the propagation environment. Hence the angular domain channels $\{{\bf \tilde h}_k\}_{k=1}^K$ have the common sparsity\cite{WC}, namely
\begin{equation}
\mathcal{S}={\rm supp} \{ {\bf{\tilde{h}}}_1 \} = {\rm supp} \{ {\bf{\tilde{h}}}_2 \} \cdots={\rm supp} \{ {\bf{\tilde{h}}}_K \}.
\label{commonSupport}
\end{equation}
Given the meamsurement \eqref{measurement}, the channel can be acquired by solving the following optimization problem
\begin{equation}
\begin{aligned}
&\min_{{\bf{\tilde{h}}}^{\rm opt}_k,1\leq k\leq K} \Vert {\bf{\tilde{h}}}_k \Vert_0\\
&{\rm s.t.} \: \Vert {\bf{\Phi \Psi}} {\bf{\tilde{h}}}_k-{\bf{y}}_k \Vert_2 \leq \epsilon,\forall k\: {\mathrm{and}}\: \eqref{commonSupport}. 
\label{opt}
\end{aligned}
\end{equation}
To fully utilize the common sparsity feature, the greedy algorithm simultaneous orthogonal match pursuit (SOMP) is used to solve the optimization problem \eqref{opt} \cite{WC} for jointly acquiring multiple sparse channel vectors at different subcarriers.
\subsection{DL-Assisted Estimation Enhancement Architecture}
\begin{figure*}[t]
\centering
\includegraphics[scale=0.65]{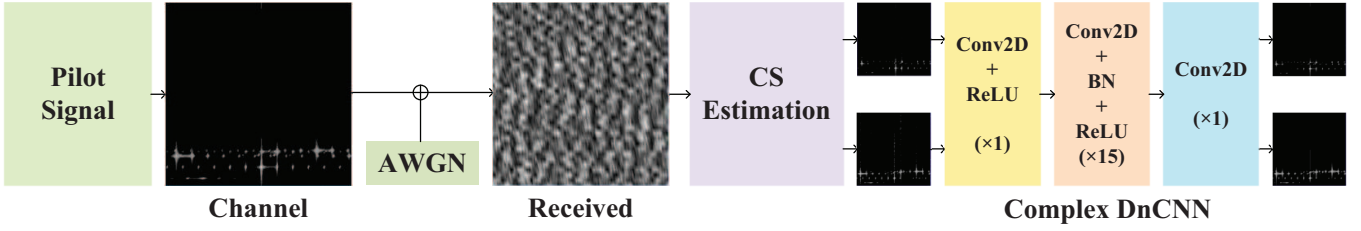}
\caption{After preliminary estimation, the CV-DnCNN model jointly processes the real part and the imaginary part of the estimated channel, and output the residual noise after pilot training.}
\label{fig:neuMod}

\end{figure*}
In color image processing tasks, a color image can be divided into several highly correlated channels (e.g., in RGB color model, color image can be divided into 3 channels, namely red, green, and blue).
Inspired by the property that the elements of the mmWave MIMO channel matrix have high correlation between the real part and imaginary part in angular-delay domain, we consider to treat this matrix as a 2-channel noisy image, so that the state-of-the-art denoising neural networks can be used to this matrix for improved estimation accuracy as shown in Fig.\ref{fig:neuMod}. 
Specifically, given the CS estimation $\{{{\bf \tilde h}_k^{\star}}\}_{k=1}^K$, the angular-delay domain MIMO channel matrix can be obtained as
\begin{equation}
{\bf{\hat{G}}} = {\bf\tilde H}^{\star} {\bf{{T}}}^{H}={\bf{{G}}} + {\bf{{E}}},
\end{equation}
where ${\bf\tilde H}^{\star}=[{\bf{{\tilde h}}}^{\star}_1,\cdots,{\bf{{\tilde h}}}^{\star}_K]  \in \mathbb{C}^{\beta^4 N_{\rm IRS}^{\rm ant}N_{\rm UE}^{\rm ant}\times K}$ is the angular-frequency channel reconstructed by SOMP algorithm, $\bf T$ denotes the $K\times K$ DFT matrix, ${\bf{{E}}}$ denotes the estimation error that can hardly be modeled, and ${\bf G}$ is the true angular-delay domain MIMO channel matrix.

There have been some denoiser-based channel estimators proposed for enhanced performance\cite{LDAMP}, where the real-part and imaginary-part of complex-valued data are processed independently. However, the operation of complex numbers should conform to their corresponding calculation rules, so the network structure needs to be modified. The proposed CV-DnCNN adopts the same network architecture as that in DnCNN\cite{DnCNN}, except for the complex signal processing modules, which can jointly process the real part and imaginary part of angular-delay domain channel matrix by exploiting their correlation for enhanced performance. More specifically, we integrate the complex building blocks inspired from\cite{cDNN} into the DnCNN, so that the denoiser can be tailored for complex signal processing. The CV-DnCNN consists of 15 repeated convolutional layers, an input convolutional layer, and an output convolutional layer. For the first layer, 64 filters of size $3\times 3$ are used to generate 64 feature maps, and rectified linear unit (ReLU) is used for activation. For the repeated 15 layers, 64 filters of size $3\times 3\times 64$ are used, and a batch normalization is adopted between convolution and ReLU to speed up the training process and improve the denoising performance. For the output layer, 1 filter of size $3\times 3\times 64$ is used for channel matrix reconstruction.
Different from conventional DnCNN, the CV-DnCNN adopts the complex convolutional layer \cite{cDNN}, which can be mathematically expressed as
\begin{align}
\left[\begin{matrix}\mathfrak{Re}({\bf W}\circledast{\bf h})\\\mathfrak{Im}({\bf W}\circledast{\bf h})\end{matrix}\right]=\left[\begin{matrix}{\bf A}& -{\bf B}\\{\bf B}& {\bf A}\end{matrix}\right]\circledast \left[\begin{matrix}{\bf x}\\{\bf y}\end{matrix}\right],
\end{align}
where ${\bf W}={\bf A}+{\bf B}j$ denotes the complex weights (filter) of convolutional networks, and ${\bf h}={\bf x}+{\bf y}j$ denotes the complex data inputed. The activation function is replaced by complex ReLU (cReLU) function\cite{cDNN}, which can be described as
\begin{align}
{\rm cReLU}({\bf h})={\rm ReLU}({\bf{x}})+{\rm ReLU}({\bf{y}})j.
\end{align}
Rebuilt by the complex building blocks, deep complex networks are verified to possess richer representational capacities\cite{cDNN}. Similar to DnCNN, the mean square error (MSE) is adopted here as loss function
\begin{equation}
\ell({\bf{\Omega}})=\frac{1}{2N}\sum_{i=1}^N\Vert \mathcal{F}({\bf{\hat{G}}}^{(i)};{\bf{\Omega}} )-({\bf{\hat{G}}}^{(i)}-{\bf{{G}}}^{(i)}) \Vert_F^2,
\end{equation}
where ${\bf{\Omega}}$ denotes the parameters in our denoising network, and the superscript $(i)$ denotes the input training data index. The DnCNN practically learns the common features among different samples, and trains the parameters of the filters (i.e., the convolutional kernels). In this way, DnCNN can extract the features of the input samples and effectively reconstruct the denoised samples. By minimizing the loss function, our purpose is to learn the mapping $\mathcal{F}$, from noisy estimated channel ${\bf{\hat{G}}}$ to noise ${\bf{{E}}}$. In other words, we can utilize the DL method to learn the residual mapping $\mathcal{F}({\bf{\hat{G}}} ) \approx {\bf{{E}}}$.
To avoid the challenging training samples acquisition and online training process, we consider to train the CV-DnCNN offline\cite{QTcomm} with the simulated channel dataset generated according to the classical geometric channel model \eqref{gm}. The feasibility of this approach and the robustness of the pretrained model will be verified in Section IV.
\begin{figure}[t]
  \centering
{\includegraphics[scale=0.55]{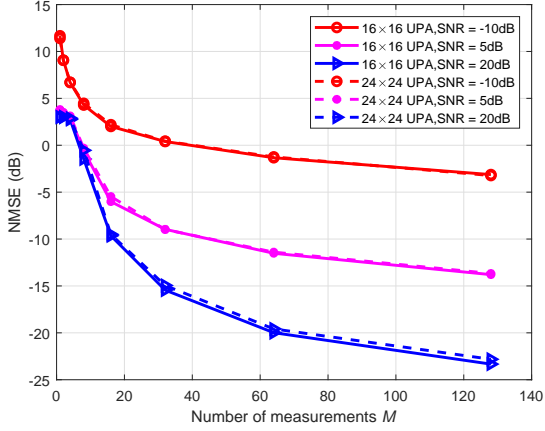}\label{fig:nmse1}}
  \caption{NMSE performance comparison of the proposed CS channel estimation scheme versus the number of active elements (i.e., the number of measurements $M$).}
 \label{fig3}
\end{figure}
\begin{figure*}[t]
  \centering
  \subfigure[]{\includegraphics[scale=0.5]{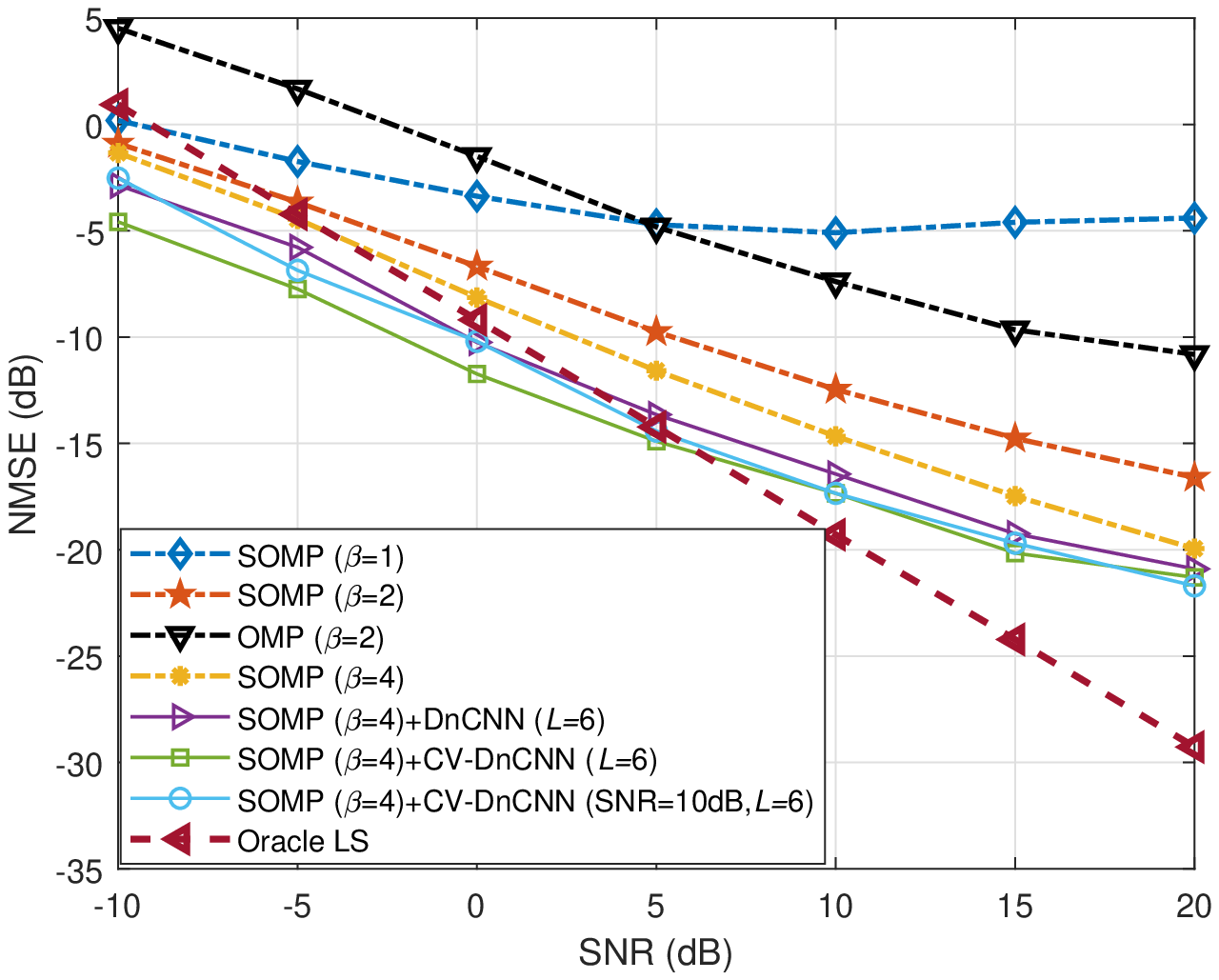}\label{fig2:nmse}}
  \subfigure[]{\includegraphics[scale=0.5]{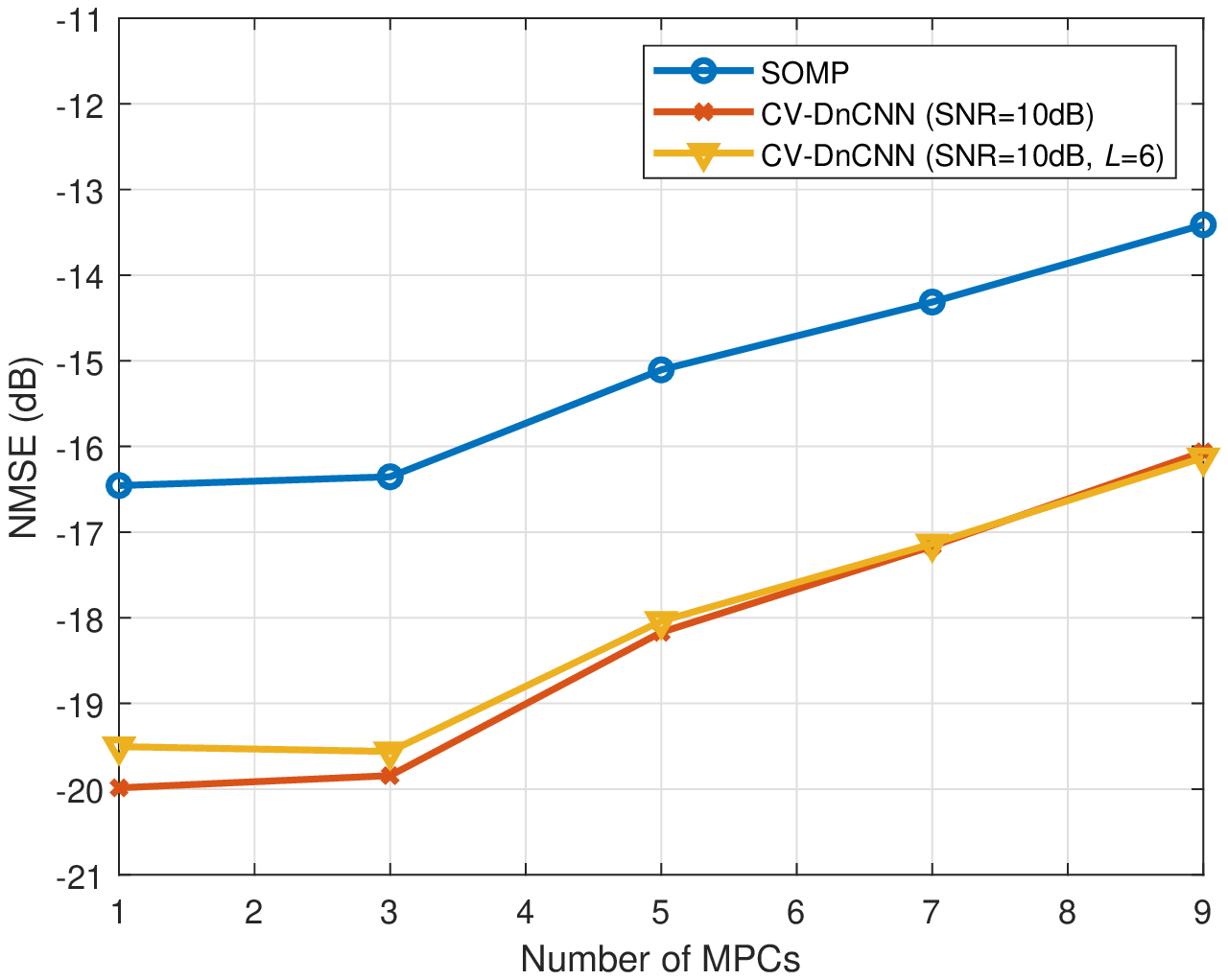}\label{NMSEvsMPC}}
  \caption{Robustness and generalization of the proposed CV-DnCNN: (a) CV-DnCNN pretrained with ${\rm SNR} = 10$ dB and $L=6$ is robust to various SNRs, (b) CV-DnCNN pretrained with  ${\rm SNR} = 10 $dB and $L=6$ can be applied to channels with various numbers of MPCs (${\rm SNR}=10$dB).}
\end{figure*}

After training the proposed CV-DnCNN, the neural network learns the mapping from noisy channel ${\bf{\hat{G}}}$ to the overall estimation error ${\bf{{E}}}$ as ${\bf{{\hat E}}} = \mathcal{F}({\bf{\hat{G}}})$, and the enhanced estimation result can be obtained as
\begin{align}
{\bf{{G}}}^{\rm e}  = {\bf{\hat{G}}}-\mathcal{F}({\bf{\hat{G}}})={\bf{{G}}} + {\bf{{E}}}-{\bf{\hat{E}}}.
\label{result}
\end{align}
Based on \eqref{result}, we can obtain the final spatial-frequency domain channel by ${\bf{\hat{H}}} =  {\bf{{G}}}^{\rm e}{\bf{{T}}}$.

\section{Simulation Results}

In this section, we investigate the performance of the proposed CV-DnCNN assisted CS channel estimation scheme with the metric of normalized MSE (NMSE) as
\begin{align}
{\rm NMSE}=\mathbb{E} \left[\frac{\Vert {\bf{H}}-{\bf {\hat H}} \Vert_{\rm{F}}^2}{{\Vert {\bf{H}} \Vert_{\rm{F}}^2}}\right],
\end{align}
where $\mathbb{E}\left[\cdot \right]$ denotes the expectation operation. In our simulations, IRS is an $N_{\rm{IRS}}^{\rm ant}=N_{\rm{IRS}}^{\rm H}\times N_{\rm{IRS}}^{\rm W}$ UPA with $N_{\rm IRS}^{\rm S}=1$ RF chain, and the UE is equipped with $N_{\rm UE}^{\rm ant}=1$ antenna, unless otherwise stated. The carrier frequency is $28{\rm{GHz}}$, and the bandwidth is $f_{\rm BW}=100$MHz with the number of OFDM's subcarriers $K=256$ in the pilot training phase. As the maximum multipath delay is limited to $\tau_{max}={32}/{f_{\rm BW}}$ in order to mitigate the effect of multipath fading, a cyclic prefix (CP) is set to $L_{\rm CP}=32$. We consider $L=6$ unless otherwise stated, and the azimuth/elevation AoAs and AoDs follow the uniform distribution $\mathcal{U}[-\pi/2,\pi/2]$. The SNR is defined by ${\rm SNR}=\mathbb{E}\{\Vert {\bf \Phi}{\bf h}\Vert_F^2/\Vert {\bf n} \Vert_F^2\}$ in \eqref{Y}. Additionally, in our deep learning experiments, we adopt $N_{\rm IRS}^{\rm ant}=16\times 16$ UPA with $M=64$ active antennas, and the SOMP oversampling rate $\beta=4$. The weights of neural network is optimized by Adam optimizer, and is fed with $5000$ training samples in training process. Training process lasts for $150$ epochs, and the batch size is set to $8$ for better convergence. Initial learning rate is set to ${\rm lr}=1\times 10^{-4}$, and decreases at epoch milestones $[30,60,90,120]$ with decreasing rate $0.4$ (i.e., learning rate decreases to $0.4$ times of last epoch when reaching the epoch milestones).

The feasibility of our adopted preliminary estimation achieved by SOMP algorithm is first investigated. Fig. \ref{fig3} shows the NMSE performance versus the number of measurements $M$. We choose the UPAs with the sizes of $16\times 16$ and $24\times 24$ for comparison, and the result obviously show that as $M$ increases, preliminary estimation achieves a better performance. Moreover, it can be observed that given $L$ and $M$, the achieved NMSE performance degrades slightly as the number of IRS elements becomes large. According to the comparison between two UPAs, it is $M$ that really affects the estimation performance rather than the proportion of active elements $M/N^{\rm ant}_{\rm IRS}$. This indicates the proposed scheme is suitable for IRS with very large aperture owing to the mmWave channel sparsity.
Considering the trade-off between complexity and the performance affected by the number of measurements $M$, we pick $M=64$ for the following simulations.


After the preliminary SOMP-based estimation with $\beta=4$, we enhance the accuracy by proposed CV-DnCNN, and the simulation results are shown in Fig. \ref{fig2:nmse}. Compared with the preliminary estimation, our proposed CV-DnCNN shows a performance gain around $4$dB, and also demonstrates the superiority beyond real-valued DnCNN.

Additionally, the robustness of CV-DnCNN also needs to be investigated. We consider to test the trained CV-DnCNN with the channel dataset, whose parameters are different from the channel dataset in the training phase. In Fig. 4(a), we apply the CV-DnCNN trained at ${\rm SNR}=10$dB in the SNR regime of $-10\sim 20\rm dB$, and the results show its good robustness. Further, we apply the CV-DnCNN pretrained at ${\rm SNR}=10$dB with $L=6$ MPCs to the channels with various numbers of MPCs as shown in Fig. \ref{NMSEvsMPC}. The pretrained CV-DnCNN can well adapt to the channels with different numbers of MPCs, and shows almost the same performance as the CV-DnCNN trained by the channel dataset with the matched SNRs and MPCs. These observations imply that the proposed CV-DnCNN possesses a good generalization ability when the parameters of training channel dataset and test channel dataset are not well matched. Therefore, in practice, the CV-DnCNN can be first offline pretrained by a simulated channel dataset, and then work in practical scenarios with negligible performance loss.

It costs less than $1\times 10^{-5}$s for CV-DnCNN to process the preliminary estimation result with an i7-8700 processor and NVIDIA 1060ti GPU per channel.

\section{Conclusion}
In this work we have adopted a passive/active IRS architecture with very few RF chains, and proposed a low training overhead CS-based broadband channel estimation method assisted with complex-valued denoising network for mmWave IRS communication systems. Our method has demonstrated that a considerable NMSE performance can be achieved with a small number of elements activated in each training phase, and the performance can be further improved by the proposed DL method within very short time. In particular, our proposed CV-DnCNN shows richer representational capacities beyond real-valued ones which results in superior performance. Moreover, the robustness makes it possible for our model to be applied in different SNR scenarios without repetitive training.

\end{document}